\documentclass[aps,twocolumn,noshowpacs,tightenlines,nobalancelastpage]{revtex4}

\usepackage{amsmath}
\usepackage{amssymb}
\usepackage{graphicx}
\usepackage{dcolumn}
\usepackage{bm}

\begin{document}

\title{State-Insensitive Cooling and Trapping of Single Atoms in an Optical Cavity}
\author{J.~McKeever, J.~R.~Buck, A.~D.~Boozer, A.~Kuzmich$^{\ast }$, H.-C.~N\"{a}gerl$^{\dagger }$, D.~M.~Stamper-Kurn$^{\ddagger }$, and H.~J.~Kimble}
\affiliation{Norman Bridge Laboratory of Physics 12-33\\
California Institute of Technology, Pasadena, CA 91125}

\begin{abstract}
Single Cesium atoms are cooled and trapped inside a small optical
cavity by way of a novel far-off-resonance dipole-force trap
(FORT), with observed lifetimes of $2-3$ seconds. Trapped atoms
are observed continuously via transmission of a strongly coupled
probe beam, with individual events lasting $\simeq 1$ s. The loss
of successive atoms from the trap $N\geq 3\rightarrow 2\rightarrow
1\rightarrow 0$ is thereby monitored in real time. Trapping,
cooling, and interactions with strong coupling are enabled by the
FORT potential, for which the center-of-mass motion is only weakly
dependent on the atom's internal state.
\end{abstract}


\maketitle

A long-standing ambition in the field of cavity quantum
electrodynamics (QED) has been to trap single atoms inside
high-$Q$ cavities in a regime of strong coupling \cite{review98}.
Diverse avenues have been pursued for creating the trapping
potential for atom confinement, including additional far
off-resonant trapping beams \cite{ye99}, near-resonant light with
with $\bar{n}\simeq 1$ intracavity photons \cite{hood00,pinkse00},
and single trapped ions in high-finesse optical cavities
\cite{walther01,blatt01}, although strong coupling has yet to be
achieved for trapped ions. A critical aspect of this research is
the development of techniques for atom localization that are
compatible with strong coupling, as required for quantum
computation and communication
\cite{pellizari95,cirac97,vanenk98,cabrillo99,bose99,parkins99}.

In this Letter we present experiments to enable quantum
information processing in cavity QED by (1) achieving extended
trapping times for single atoms in a cavity while still
maintaining strong coupling, (2) realizing a trapping potential
for the center-of-mass motion that is largely independent of the
internal atomic state, and (3) demonstrating a scheme that allows
continuous observation of trapped atoms by way of the atom-field
coupling. More specifically, we have recorded trapping times up to
$3~\mathrm{s}$ for single Cs atoms stored in an intracavity
far-off resonance trap (FORT) \cite{metcalf99}, which represents
an improvement by a factor of $10^{2}$ beyond the first
realization of trapping in cavity QED \cite{ye99}, and by roughly
$10^{4}$ beyond prior results for atomic trapping \cite{hood00}
and localization \cite{pinkse00} with $\bar{n}\simeq 1$ photon. We
have also continuously monitored trapped atoms by way of strong
coupling to a probe beam, including observations of trap loss atom
by atom over intervals $\simeq 1$ s. These measurements
incorporate auxiliary cooling beams, and provide the first
realization of cooling for trapped atoms strongly coupled to a
cavity. Our protocols are facilitated by the choice of a
\textquotedblleft magic\textquotedblright\ wavelength for the FORT
\cite{hood99,katori99,vanenk00}, for which the relevant atomic
levels are shifted almost equally, thereby providing significant
advantages for coherent state manipulation of the atom-cavity
system.

\begin{figure}[t]
\includegraphics[width=8.6cm]{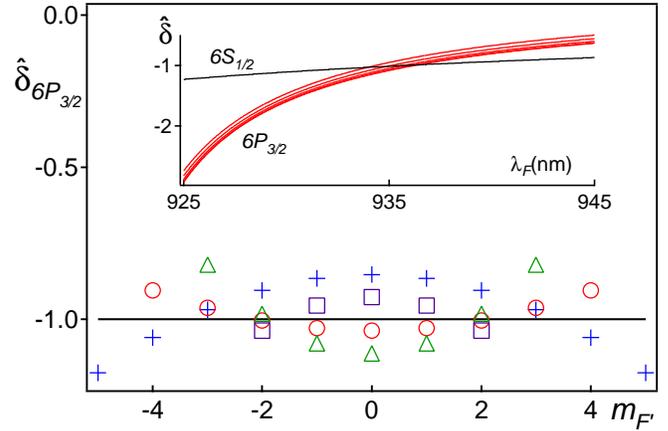}
\caption{\label{starkfig} AC-Stark shifts
$(\hat{\delta}_{6S_{1/2}},\hat{\delta}_{6P_{3/2}})$ for the
$(6S_{1/2},6P_{3/2})$ levels in atomic Cs for a linearly polarized
FORT. The inset shows
$(\hat{\delta}_{6S_{1/2}},\hat{\delta}_{6P_{3/2},F^{\prime }=4})$
as functions of FORT wavelength $\lambda_{F}$. The full plot gives
$\hat{\delta}_{6P_{3/2}}$ versus $m_{F^{\prime }}$ for each of the
levels $6P_{3/2},F^{\prime }=2,3,4,5$ for
$\lambda_{F}=935.6~\mathrm{nm}$. In each case, the normalization
is $\hat{\delta}=\delta /[\delta
_{6S_{1/2}}(\lambda_{F}=935.6~\text{nm})]$ \cite{FORT-details}.}
\end{figure}

A major obstacle to the integration of a conventional red-detuned
FORT within the setting of cavity QED is that excited electronic
states generally experience a positive AC-Stark shift of
comparable magnitude to the negative (trapping) shift of the
ground state \cite{metcalf99}. This leads to the unfortunate
consequence that the detuning and hence the effective coupling
between an atomic transition and the cavity mode become strong
functions of the atom's position within the trap \cite{vanenk00}.
However, due to the specific multi-level structure of Cesium, the
wavelength $\lambda_{F}$ of the trapping laser can be tuned to a
region where both of these problems are eliminated for the
$6S_{1/2}\rightarrow 6P_{3/2}$ transition, as illustrated in
Fig.~\ref{starkfig} \cite{hood99,katori99,vanenk00,FORT-details}.
Around the \textquotedblleft magic\textquotedblright\ wavelength
$\lambda_{F}=935~\mathrm{nm}$, the sum of AC-Stark shifts coming
from different allowed optical transitions results in the ground
$6S_{1/2}$ and excited $6P_{3/2}$ states both being shifted
downwards by comparable amounts, $\delta _{6S_{1/2}}\simeq
\delta_{6P_{3/2}}$, albeit with small dependence on
$(F^{\prime},m_{F^{\prime }})$ for the shifts $\delta_{6P_{3/2}}$.

\begin{figure}[tb]
\includegraphics[width=8.6cm]{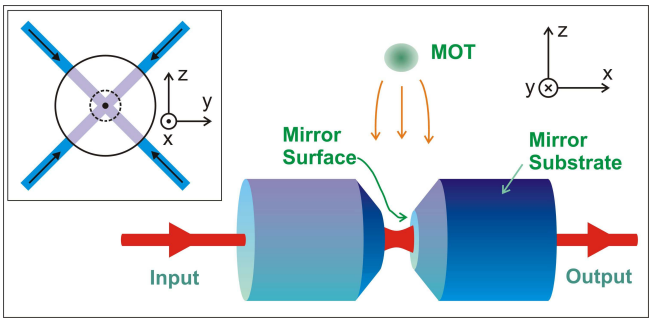}
\caption{Schematic of experiment for trapping single atoms in an
optical cavity in a regime of strong coupling. Relevant cavity
parameters are length $l=43.0~\mathrm{\mu m}$, waist
$w_{0}=23.9~\mathrm{\mu m}$, and finesse $\mathcal{F}=4.2\times
10^{5}$ at $852~\mathrm{nm}$. The inset illustrates transverse
beams used for cooling and repumping.} \label{setup}
\end{figure}

The task then is to achieve state-independent trapping while still
maintaining strong coupling for the $6S_{1/2}\rightarrow 6P_{3/2}$
transition. Our experimental setup to achieve this end is
schematically depicted in Fig.~\ref{setup} \cite{ye99}.
Significantly, the cavity has a $\mathrm{TEM}_{00}$ longitudinal
mode located nine mode orders below the mode employed for cavity
QED at $852~\mathrm{nm}$, at the wavelength
$\bar{\lambda}_{F}=935.6~\mathrm{nm}$, allowing the implementation
of a FORT with $\delta _{6S_{1/2}}\simeq \delta_{6P_{3/2}}$. The
field to excite this cavity mode is provided by a laser at
$\bar{\lambda}_{F}$, which is independently locked to the cavity.
The finesse of the cavity at $\bar{\lambda}_{F}$ is
$\mathcal{F}\sim 2200$ \cite{hood01}, so that a mode-matched input
power of $1.2~\mathrm{mW}$ gives a peak AC-Stark shift
$\delta_{6S_{1/2}}/2\pi=-47~\mathrm{MHz}$ for all states in the
$6S_{1/2}$ ground manifold, corresponding to a trap depth
$U_{0}/k_{B}=2.3~\mathrm{mK}$, which was used for all experiments.

Principal parameters relevant to cavity QED with the system in Fig.~%
\ref{setup} are the Rabi frequency $2g_{0}$ for a single quantum
of excitation and the amplitude decay rates $(\kappa ,\gamma )$
due to cavity
losses and atomic spontaneous emission. For our system, $g_{0}/2\pi =24~%
\mathrm{MHz}$, $\kappa /2\pi =4.2~\mathrm{MHz}$, and $\gamma /2\pi =2.6~%
\mathrm{MHz}$, where $g_{0}$ is for the
$(6S_{1/2},F=4,m_{F}=4)\rightarrow
(6P_{3/2},F^{\prime }=5,m_{F}^{\prime }=4)$ transition in atomic Cs at $%
\lambda _{0}=852.4~\mathrm{nm}$. Strong coupling is thereby achieved $%
(g_{0}\gg (\kappa ,\gamma ))$, resulting in critical photon and
atom numbers $n_{0}\equiv \gamma ^{2}/(2g_{0}^{2})\simeq 0.006$,
$N_{0}\equiv 2\kappa \gamma /g_{0}^{2}\simeq 0.04$. The small
transition shifts for our FORT mean that $g_{0}$ is considerably
larger than the spatially dependent shift $\delta_{0}$ of the bare
atomic frequency employed for cavity QED, $g_{0}\gg \delta
_{0}\equiv |\delta _{6P_{3/2}}-\delta _{6S_{1/2}}|$, whereas in a
conventional FORT, $\delta _{0}\sim 2|\delta _{6S_{1/2}}|\gg
g_{0}$.

In addition to the FORT field, the input to the cavity consists of
probe and locking beams, all of which are directed to separate
detectors at the output. The transmitted probe beam is monitored
using heterodyne detection, allowing real-time detection of
individual cold atoms within the cavity mode \cite{mabuchi96}. The
cavity length is actively controlled using a cavity resonance at
$\lambda_{C}=835.8~\mathrm{nm}$, so the length is stabilized and
tunable independently of all other intracavity fields \cite{ye99}.
The probe as well as the FORT beam are linearly polarized along a
direction $\hat{l}_{+}$ orthogonal to the $x$-axis of the cavity
\cite{hood01,birefringence}.

Cold atoms are collected in a magneto-optical trap (MOT) roughly
$5~\mathrm{mm}$ above the cavity mirrors and then released after a
stage of sub-Doppler polarization-gradient cooling
\cite{metcalf99}. Freely falling atoms arrive at the cavity mode
over an interval of about $10~\mathrm{ms}$, with kinetic energy
$E_{K}/k_{B}\simeq 0.8~\mathrm{mK}$, velocity $v\simeq
0.30~\mathrm{m/s}$, and transit time $\Delta t=2w_{0}/v\simeq
150~\mathrm{\mu s}$. Two additional orthogonal pairs of
counter-propagating beams in a $\sigma^{+}-\sigma^{-}$
configuration illuminate the region between the cavity mirrors
along directions at $\pm 45^{\circ }$ relative to
$\hat{y},\hat{z}$ (the ``$y-z$ beams'') and contain cooling light
tuned red of $F=4\rightarrow F^{\prime }=5$ and repumping light
near the $F=3\rightarrow F^{\prime }=3 $ transition
\cite{intensity}. These beams eliminate the free-fall velocity to
capture atoms in the FORT and provide for subsequent cooling of
trapped atoms.

We employed two distinct protocols to study the lifetime for
single trapped atoms in our FORT.

\textbf{(1)} \textit{Trapping \textquotedblleft in the
dark\textquotedblright} with the atom illuminated only by the FORT
laser at $\bar{\lambda}_{F}$ and the cavity-locking laser at
$\lambda_{C}$. For this protocol, strong coupling enables
real-time monitoring of single atoms within the cavity for initial
triggering of cooling light and for final detection.

\textbf{(2)} \textit{Trapping with continuous observation of
single atoms} with cavity probe and cooling light during the
trapping interval. In this case, atoms in the cavity mode are
monitored by way of the cavity probe beam, with cooling provided
by the auxiliary $y-z$ beams.

\textbf{(1)} In our first protocol, the $F=4\rightarrow F^{\prime
}=5$ transition is strongly coupled to the cavity field, with zero
detuning of the cavity from the bare atomic resonance,
$\Delta_{C}\equiv\omega_{C}-\omega_{4\rightarrow 5}=0$. In
contrast to Ref.~\cite{ye99}, here the FORT is \textit{ON}
continuously without switching, which makes a cooling mechanism
necessary to load atoms into the trap. The initial detection of a
single atom falling into the cavity mode is performed with the
probe beam tuned to the lower sideband of the vacuum-Rabi spectrum
($\Delta_{p}=\omega _{p}-\omega _{4\rightarrow 5}=-2\pi \times
20~\mathrm{MHz}$). The resulting \textit{increase} in transmitted
probe power when an atom approaches a region of optimal coupling
\cite{hood98,trigger} triggers \textit{ON} a pulse of transverse
cooling light from the $y-z$ beams, detuned $41~\mathrm{MHz}$ red
of $\omega_{4\rightarrow 5}$. During the subsequent trapping
interval, all near-resonant fields are turned \textit{OFF}
(including the transverse cooling light). After a variable delay
$t_{T}$, the probe field is switched back \textit{ON} to detect
whether the atom is still trapped, now with $\Delta_{p}=0$.

\begin{figure}[tb]
\includegraphics[width=8.6cm]{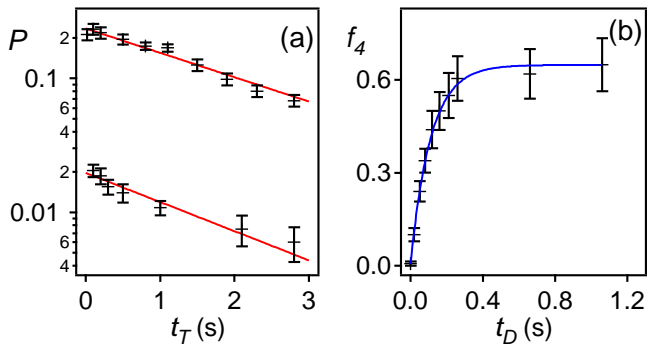}
\caption{\label{lifetime}(a) Detection probability $P$ as a
function of trapping time $t_{T}$. The upper data set is for mean
intracavity atom number $\bar{N}\approx 0.30$, while the lower set
is for $\bar{N}\approx 0.019$ atoms. Exponential fits (solid
lines) yield lifetimes $\tau_{\text{upper}}=(2.4\pm 0.2)$ s and
$\tau_{\text{lower}}=(2.0\pm 0.3)$ s. (b) The fractional
population $f_{4}(t_{D})$ in $F=4$ following depletion of this
level at $t_{D}=0$. An exponential fit (solid line) gives
$\tau_{R}=(0.11\pm 0.02)$ s.}
\end{figure}

Data collected in this manner are shown in Fig.~\ref{lifetime}(a),
which displays the conditional probability $P$ to detect an atom
given an initial single-atom triggering event versus the time
delay $t_{T}$. The two data sets shown in Fig.~\ref{lifetime}(a)
yield comparable lifetimes, the upper acquired with mean
intracavity atom number $\bar{N}=0.30$ atoms and the lower with
$\bar{N}=0.019$ \cite{nbar}. The offset in $P$ between these two
curves arises primarily from a reduction in duration $\delta t$ of
the cooling pulses, from $100~\mathrm{\mu s}$ to $5~\mathrm{\mu s
}$, which results in a reduced capture probability. Measurements
with constant $\delta t$ but with $\bar{N}$ varied by adjusting
the MOT parameters allow us to investigate the probability of
trapping an atom other than the ``trigger'' atom and of capturing
more than one atom. For example, with $\delta t =5~\mathrm{\mu s}$
as in the lower set, we have varied $0.011\lesssim \bar{N}\lesssim
0.20$ with no observable change in either $P_{T}$ or the trap
lifetime $\tau $. Since a conservative upper bound on the relative
probability of trapping a second atom is just $\bar{N}/2$ (when
$\bar{N}$ $\ll 1$), these data strongly support the conclusion
that our measurements are for single trapped atoms. We routinely
observe lifetimes $2~\mathrm{s}<\tau <3~\mathrm{s}$ depending upon
the parameters chosen for trap loading and cooling.

Fig.~\ref{lifetime}(b) explores scattering processes within the
FORT that transfer population between the $6S_{1/2},F=(3,4)$
ground-state hyperfine levels. For these measurements, the $F=4$
level is initially depleted, and then the population in $F=4$ as
well as the total $3+4$ population are monitored as functions of
time $t_{D}$ to yield the fractional population $f_{4}(t_{D})$ in
$F=4$. The measured time $\tau_{R}=(0.11\pm 0.02)$s for
re-equilibration of populations between $F=(3,4)$ agrees with a
numerical simulation based upon scattering rates in our FORT,
which predicts $\tau_{R}=0.10~\mathrm{s}$ for atoms trapped at the
peak FORT intensity in an initially unpolarized state in the $F=3$
level.

Turning next to the question of the mechanisms that limit our FORT
lifetime, we recall that parametric heating caused by intensity
fluctuations of the trapping field can be quite important
\cite{ye99,savard97}. From measurements of intensity fluctuations
for our FORT around twice the relevant harmonic frequencies $(\nu
_{axial}=570,\nu _{radial}=4.8)~\mathrm{kHz}$, we estimate a lower
bound to the FORT lifetime of $\tau _{p}^{axial}>1.6~\mathrm{s}$
\cite{radial}. Since this estimate
suggests that parametric heating could be a limiting factor in Fig.~\ref%
{lifetime}, we performed subsequent measurements in which the
intensity noise was reduced below the shot-noise level of our
detection system, giving a lower bound $\tau
_{p}^{axial}>9~\mathrm{s}$. Unfortunately, the measured FORT
lifetime increased only modestly to $\tau =(3.1\pm
0.4)~\mathrm{s}$, indicating that other mechanisms are partially
responsible for the observed decay.

A second suspect is a heating process described by Corwin
\textit{et al}. \cite{corwin99} associated with inelastic Raman
scattering in an elliptically polarized FORT field
\cite{birefringence}. We calculate rates $\Gamma_{s}$ for
spontaneous Raman scattering in our FORT to be $2.5$ to $7$
$\mathrm{s}^{-1}$ for transitions that change the hyperfine
quantum number $F$, and between $0.8$ and $2.5$ $\mathrm{s}^{-1}$
when only $m_{F}$ changes \cite{cline94}. Based on Eq.~3 in
Ref.~\cite{corwin99} (a two-state model), we estimate an upper
limit to the heating rate from this mechanism,
$\Gamma_{IR}\lesssim 0.2\Gamma_{s}$, giving heating times as short
as $0.7~\mathrm{s}$ for the fastest calculated scattering rate.
However, we have also undertaken a full multilevel simulation of
the optical pumping processes, which indicates much slower
heating, $\Gamma_{IR}\sim 0.02~\mathrm{s}^{-1}$. We are working to
resolve this discrepancy.

A third suspect that cannot be discounted is the presence of stray
light, which we have endeavored to eliminate. For lifetimes as in
Fig.~\ref{lifetime}, we require intracavity photon number
$\bar{n}\ll 10^{-5}$, which is not trivial to diagnose. A final
concern is the background pressure in the region of the FORT.
Although the chamber pressure is $3\times 10^{-10}~\mathrm{Torr}$
(leading to $\tau\simeq 30~\mathrm{s}$), we have no direct
measurement of the residual gas density in the narrow cylinder
between the mirror substrates (diameter $1~\mathrm{mm}$ and length
$43~\mathrm{\mu m}$), except for the trap lifetime itself.

\textbf{(2)} Toward the goals of continuous observation of single
trapped atoms \cite{hood00,pinkse00} and of implementing $\Lambda
$-schemes in cavity QED
\cite{pellizari95,cirac97,vanenk98,kuhn02}, we next present
results from our second protocol. Here, the $F=4\rightarrow
F^{\prime }=4$ transition is strongly coupled to the cavity field,
with $\Delta _{C}^{\prime }\equiv \omega _{C}-\omega
_{4\rightarrow 4}=0$. In contrast to our protocol \textbf{(1)},
the FORT and the transverse $y-z$ beams are left \textit{ON}
continuously, with the latter containing only light near the
$F=3\rightarrow F^{\prime }=3$ resonance, with detuning $\Delta
_{3}$. Significantly, we observe trap loading with \textit{no
cooling light near the $F=4\rightarrow F^{\prime }=5$ transition}.

\begin{figure}[tb]
\includegraphics[width=8.6cm]{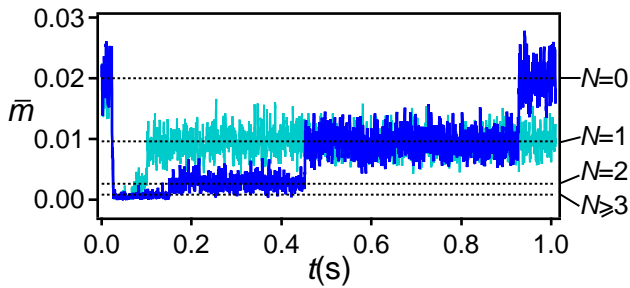}
\caption{\label{cwprobe} Two traces of the continuous observation
of trapped atoms inside a cavity in a regime of strong coupling.
After an initial sharp reduction around $t=0$ as atoms are cooled
into the cavity mode, the intracavity field strength $\bar{m}$
increases in a discontinuous fashion as trapped atoms escape from
the cavity mode one by one. RF detection bandwidth
$=1~\mathrm{kHz}$, $\Delta_{C}^{\prime }=0=\Delta_{p}^{\prime }$,
and $\Delta_{3}/2\pi =25~\mathrm{MHz}$ (\textit{blue}).}
\end{figure}

An example of the resulting probe transmission is shown in
Fig.~\ref{cwprobe}, which displays two separate records of the
continuous observation of trapped atoms. Here, the probe detuning
$\Delta _{p}^{\prime }=\omega _{p}-\omega _{4\rightarrow 4}=0$ and
the probe strength is given in terms of $\bar{m}=|\langle
\hat{a}\rangle |^{2}$ deduced from the heterodyne current, with
$\hat{a}$ as the annihilation operator for the intracavity field.
We believe that the $y-z$ repumping beams (which excite
$F=3\rightarrow F^{\prime }=3$) provide cooling, since without
them the atoms would ``roll'' in and out of the near-conservative
FORT potential (indeed no trapping occurs in their absence). In
addition, this is a continuous cooling and loading scheme, so that
we routinely load multiple atoms into the trap.

The most striking characteristic of the data collected in this
manner is that $\bar{m}$ versus $t$ always reaches its deepest
level within the $\simeq 10$ ms window when the falling atoms
arrive, subsequently increasing in a discontinuous ``staircase''
of steps. As indicated in Fig.~\ref{cwprobe}, our interpretation
is that there is a different level for $\bar{m}$ associated with
each value $N$ of the number of trapped atoms (with the level
decreasing for higher $N$), and that each step is due to the loss
of an atom from the cavity mode. In addition, we observe a strong
dependence both of the initial trapping probability and of the
continuous observation time on the detuning of the transverse
beams, with an optimal value $\Delta _{3}\simeq 25~\mathrm{MHz}$
to the \textit{blue} of the $3\rightarrow 3$ transition, which
strongly suggests blue Sisyphus cooling \cite{boiron96}.

We stress that observations as in Fig.~\ref{cwprobe} are made
possible by strong coupling in cavity QED, for which individual
intracavity atoms cause the displayed changes in probe
transmission. While $\bar{m}$ in Figure 4 is only $\simeq 0.01$,
it represents an output flux $\simeq 5\times 10^{5}$ photons per
second. The probe is also critical to the cooling, although it is
not clear whether this beam is acting as a simple
\textquotedblleft repumper\textquotedblright\ \cite{boiron96} or
is functioning in a more complex fashion due to strong coupling.
We have not seen such striking phenomena under similar conditions
for cavity QED with the $F=4\rightarrow F^{\prime }=5$ transition.
Note that our ability to monitor the atom as well as to cool its
motion are enabled by the state-insensitive character of the trap,
since the net transition shifts are small, $(g_{0},\Delta_{3})\gg
\delta_{0}$.

In summary, we have demonstrated a new set of ideas within the
setting of cavity QED, including state insensitive trapping
suitable for strong coupling. Trapping of single atoms with
$g_{0}\gg (\delta_{0},\kappa ,\gamma )$ has been achieved with
lifetimes $\tau \simeq 2-3$s. Since intrinsic heating in the FORT
is quite low ($\sim 11~\mathrm{\mu K/s}$ due to photon recoil), we
anticipate extensions to much longer lifetimes. Continuous
observations of multiple atoms in a cavity have been reported, and
involve an interplay of a strongly coupled probe field for
monitoring and a set of $y-z$ cooling beams. Our measurements
represent the first demonstration of cooling for trapped atoms
strongly coupled to a cavity. Beyond its critical role here, state
insensitive trapping should allow the application of diverse laser
cooling schemes, leading to atomic confinement in the Lamb-Dicke
regime with strong coupling, and thereby to further advances in
quantum information science.

We gratefully acknowledge the contributions of K.~Birnbaum,
A.~Boca, T.~W.~Lynn, S.~J.~van~Enk, D.~W.~Vernooy, and J.~Ye. This
work was supported by the Caltech MURI Center for Quantum Networks
under ARO Grant No. DAAD19-00-1-0374, by the National Science
Foundation, and by the Office of Naval Research.

$^{\ast }$Georgia Institute of Technology, Atlanta, GA 30332

$^{\dagger }$Institut f{\"u}r Experimentalphysik, Universit{\"a}t
Innsbruck, Technikerstra{\ss}e 25, A-6020 Innsbruck, Austria

$^{\ddagger }$Department of Physics, University of California,
Berkeley, CA 94720


\begin{thebibliography}{99}
\bibitem{birefringence}
Because of small stress-induced birefringence in the cavity
mirrors, we align the directions of linear polarization along an
axis that coincides with one of the cavity eigen-polarizations
\cite{hood01}, denoted by $\hat{l}_{\pm }$. For initial
polarization along $\hat{l}_{+}$, measurements of FORT [probe]
polarization along $\hat{l}_{-}$ for the cavity output power $P$
give $P_{-}/P_{+}<0.02[0.002]$ for the FORT [probe] beam.
\bibitem{intensity}
The (incoherent) sum of the four intensities is $I_{4-5}\sim
60$mW/cm$^{2}$ for the cooling and $I_{3-3}\sim 40$mW/cm$^{2}$ for
the repumping light, with uncertainties of roughly $2\times $.
\bibitem{trigger}
Specific examples of single-atom detection events are omitted
here. For $\Delta_{p}\simeq -g_{0}$, the increases in cavity
transmission are quite similar to those in Refs.
\cite{hood00,hood98}, while for $\Delta_{p}=0$, the decreases are
similar to those in Refs.~\cite{ye99,mabuchi96}, albeit now in the
presence of the FORT.
\bibitem{nbar}
$\bar{N}$ is estimated from the mean number of atom transit events
(of duration $\simeq 150\mu$s) during the interval $\simeq 10$ms
from the falling MOT atoms, in the absence of trapping.
\end{thebibliography}
\end{document}